# Data Comets: Designing a Visualization Tool for Analyzing Autonomous Aerial Vehicle Logs with Grounded Evaluation

David Saffo[1] Aristotelis Leventidis[1] Twinkle Jain[1] Michelle A. Borkin[1] Cody Dunne[1]

[1]Northeastern University, Khoury College of Computer Sciences

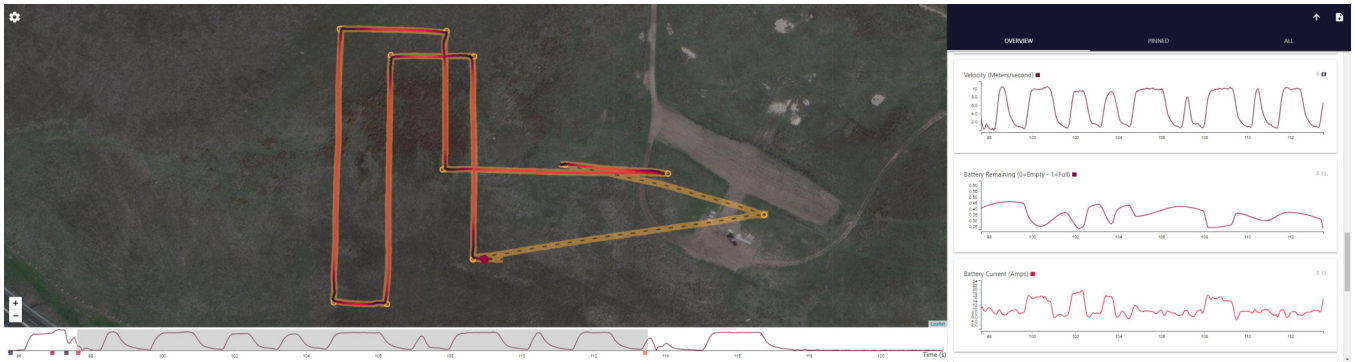

**Figure 1:** DATA COMETS *visualizes autonomous aerial vehicle logs using three main views: the map, attributes tree, and timeline. Here we see a flight log from a quadrocopter. The flight map (top-left) displays the actual flight path. To enable comparison with the programmed path, autonomous flight setpoints ○ and the predicted flight path ▬ are optionally shown. The attribute tree (right) includes a scrollable overview tab containing line charts for many commonly analyzed attributes. The velocity attribute is selected ▮ and shown on the flight path using color from slow to fast ▬▬▬. The timeline (bottom) shows a line chart of the selected attribute superimposed with colored squares for events, e.g., switches between flight modes. The timeline is brushed to filter all other views to the interval between the start ▮ and end ▮ of autonomous flight mode. The filtered portion of the flight path on the map is optionally shown as a dashed line ▬ ▬ ▬.*

**Abstract**
*Autonomous unmanned aerial vehicles are complex systems of hardware, software, and human input. Understanding this complexity is key to their development and operation. Information visualizations already exist for exploring flight logs but comprehensive analyses currently require several disparate and custom tools. This design study helps address the pain points faced by autonomous unmanned aerial vehicle developers and operators. We contribute: a spiral development process model for grounded evaluation visualization development focused on progressively broadening target user involvement and refining user goals; a demonstration of the model as part of developing a deployed and adopted visualization system; a data and task abstraction for developers and operators performing post-flight analysis of autonomous unmanned aerial vehicle logs; the design and implementation of* DATA COMETS, *an open-source and web-based interactive visualization tool for post-flight log analysis incorporating temporal, geospatial, and multivariate data; and the results of a summative evaluation of the visualization system and our abstractions based on in-the-wild usage. A free copy of this paper and source code are available at* osf.io/h4p7g

## 1. Introduction

Unmanned aerial vehicles (UAVs) have proliferated in recent years, with applications in construction [HMB18b], agriculture [MD18], cinema [MMNP18], conservation [HMB*18a], and critical medical delivery [AK19]. One reason for the rapid adoption of UAVs is the advent of airframe designs such as quadrocopters that are capable of stable flight, come in many sizes, and can be programmed for autonomous operation. These developments likely informed NASA's Dragonfly mission, which plans to send a quadrocopter-like UAV to explore the surface of Saturn's moon Titan [LTB*18]. A UAV capable of autonomous flight is well suited for navigating Titan's harsh landscape and the 159-minute round trip communication latency with Earth. However, autonomous flight adds a non-trivial amount of complexity into an already complex system.

Autonomous UAV flight involves a complex system of sensors, hardware, software, and — on some level — human input. As a simpler example, imagine a quadrocopter programmed to fly autonomously and survey crops for signs of drought stress. During the five-minute flight the UAV veers off course and crashes. This incident could have been precipitated by any number of internal or external factors — low battery, miscalibrated altitude sensors, erroneous location information, autonomous software bugs, RF or cyber attack, bird strike... When the vehicle is recovered some possibilities can be eliminated, but detecting more insidious software and hardware issues will require analyzing the flight logs. However, the logs of our short five-minute flight include over 30MB of data collected from the on-board systems, even after excluding the captured mission data (e.g., video of cropland).

Information visualizations can serve as an exploratory aid for analyzing autonomous UAV flights, as the logs contain a wide variety of interrelated and well-structured factors but users may find it difficult to verbalize their information need [FvWSN08]. From our interviews and observations of autonomous UAV developers and operators, it became clear to us that flight logging architectures are already being built with visual analysis in mind, e.g., PX4 [MHP15]. Moreover, visualizations are becoming integral to the development, operation, and analysis of many autonomous UAV systems.

The visualization tools currently used for analyzing autonomous flight log data do not incorporate the state of the art in terms of visualizing geospatial movement data, and often violate visualization best practices. As a result, analysis with these tools can be inefficient or even inaccurate, and users are often required to switch between several complimentary tools in order to conduct a comprehensive analysis. However, these tools have been developed by UAV developers and operators, to the best of their ability, to meet their needs. Understanding the design of these tools can help ensure new visualization designs not only meet user tasks, but also still feel familiar. To achieve this, it is important to have a visualization design model that will guide development of tools in context of their intended use such as grounded evaluation [IZCC08].

We conducted a design study [SMM12] in order to help address the pain points autonomous UAV developers and operators face when visually analyzing flight logs. This paper is structured to discuss the contributions of our design study in order:

1. *A "spiral" model for grounded evaluation* visualization development based on progressively broadening target user involvement and refining user goals as an extension of grounded evaluation [IZCC08].
2. *An end-to-end practical demonstration of the spiral development process* from initial evaluation to deployed and adopted system.
3. *A data and task abstraction* for autonomous aerial vehicle operators and developers performing post-flight analysis.
4. *The design and implementation of* DATA COMETS, an open-source and web-based interactive visualization tool to support post-flight analysis of autonomous aerial vehicle logs including temporal, geospatial, and multivariate data.
5. *The results of a summative evaluation of* DATA COMETS *and our abstractions* based on in-the-wild usage by target users across five weeks which validates our approach.

## 2. Related Work

In the following subsections we discuss prior work that guided our research. At the end of each subsection we define a research question, motivated by the respective material, that we aim to answer.

### 2.1. Visualization Development Processes

Visualization researchers and designers often follow predefined development models or frameworks to help guide their work. We reviewed a variety of both grounded and traditional development models in an attempt to better understand the connections, similarities, and differences between these models. Development processes can be cyclic and flexible in nature, allowing the researcher to return to previous stages and iterate based on new knowledge. Many cyclic processes have been proposed pertaining to task-based visualization development, e.g., Isenberg et al. [IZCC08], Lloyd [LD11]. Several cyclic processes are shown in fig. 2 [IZCC08, MMAM14, Mun09, RCL*05, RRF*10, SMM12]. There we show the relationships between the stages in these models and compare them with our own. These models generally share the same fundamental stages of development. However, where they differ is in the ordering of the stages and in the specificity of their processes. For example, Isenberg et al. [IZCC08] introduced a grounded evaluation approach that begins the cycle at evaluation instead of task analysis and design. Lloyd and Dykes [Llo09] evaluates human centered approaches for geo-visualizations and documents several design studies starting at both of these stages.

While most approaches conduct some form of preliminary literature search, grounded evaluation is unique because an initial evaluation of domain tasks, users, data, and existing tools is done before moving on to task analysis and design. By doing this, a grounded evaluation approach allows for a more complete understanding of the domain and issues that need to be solved. One limitation of [IZCC08] is that it lacked clear specifications for what processes should occur at each stage and how they change over iterations of the cycle. We aim to provide a practical example of these processes, similar to [SMM12, MMAM14] presenting the framework or model alongside the theoretical development process that help guide users with clear next steps. Moreover, by following the connections in fig. 2 we can draw guidance from the related stages presented in these models even when not directly using them.

**Research Question:** What does a grounded evaluation model for visualization development look like, both in theory and practice?

### 2.2. Geospatial Visualization of Movement Data

Visualizations of geospatial movement data commonly utilize a space-time cube where two or three spatial dimensions are encoded along with time [AAB*13, Kra03]. These concepts are the basis for several frameworks and models proposed to describe the many different components of visualizing this data.

Peuquet [Peu94] describes geospatial movement data as comprising of three components: space (*where*), time (*when*), and objects (*what*). Pequet argues there are thus three possible types of questions concerning this data: given *when+where* find *what*, given *when+what* find *where*, and given *where+what* find *when*. Roth [Rot13] provides an interaction taxonomy for geovisualization that identifies several tasks users would want to perform. Many of the interactions discussed are aimed at answering Peuquet's three questions. This work indicates that these tasks, despite their simplicity, are a critical part of analyzing geospatial movement data.

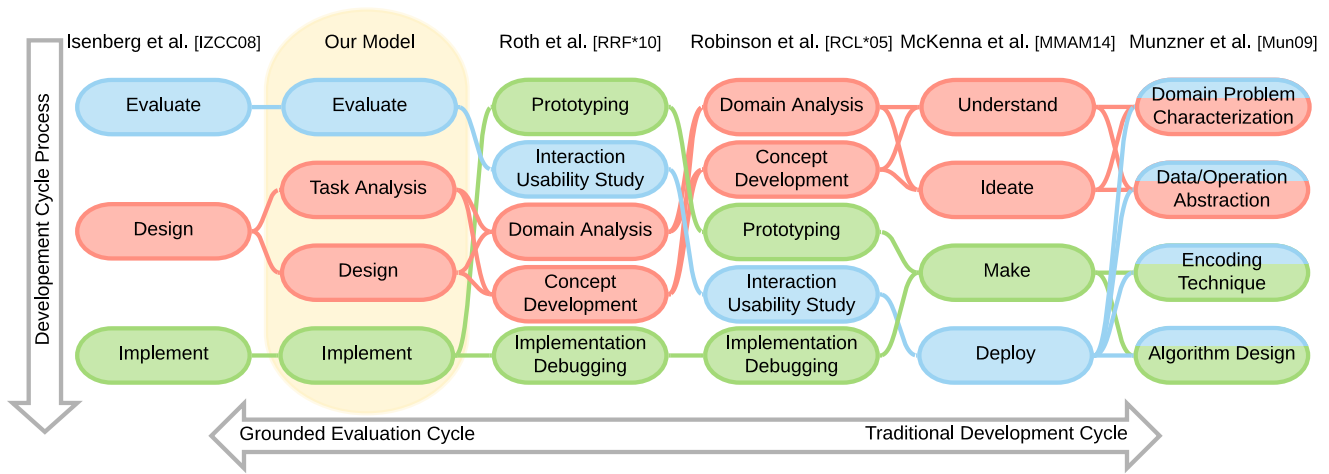

**Figure 2:** *A selection of visualization design cycle models and the relationships between their stages. The stages of each model are organized from top to bottom in their prescribed order. The cyclic aspects of these processes are not shown but implicit. Models are organized left to right according to how close they are to Isenberg et al.'s grounded evaluation model [IZCC08] or a traditional development cycle model such as Munzner et al. [Mun09]. Stages that describe similar processes are linked and categorically colored by the stages in the grounded evaluation model: Evaluate, Design, and Implement. Munzner et al. do not define a explicit evaluation stage, instead evaluation is implied within each stage as shown by their dual categorical coloring. By mapping stage relationships across models, we can more clearly see the similarities and differences in their prescribed orderings and divisions. We can leverage these similarities to find recommendations from each to aid us in development. E.g., Isenberg et al. do not provide much detail about what takes place during the Design stage. However, as this stage overlaps with McKenna et al.'s Understand and Ideate stages [MMAM14] we can draw from their recommendations.*

Bach et al. [BDA*17] present a descriptive framework based on space-time cube operations. They introduce a taxonomy that details several visualization strategies including point extraction, non-planner drilling, time chopping, time flattening, and time or space coloring. This work provides a starting point when searching for existing encoding techniques often employed for this data. We saw the potential for techniques such as time chopping, displaying only a chunk of the data at a time, to be used as an interaction method to allow our users to filter and animate the movement of their data.

Dodge et al. [DWL08] introduce a taxonomy for movement patterns important for trajectory analysis. Their taxonomy defines several patterns that can be described by their parameters such as position, distance, and speed. However, this taxonomy does not extend beyond information that can be derived from the trajectory itself. Without studying our target domain it is unclear whether these patterns are important for the analysis of a single UAV flight log.

Applications of many of the concepts and principles covered by the work cited above can be found in [AA13, AAB*13, GAA04, HCC*19] It is particularly common to encode movement as 2D or 3D paths superimposed on a map (non-planner drilling, time flattening). A slider or brushable timeline is often added to filter the displayed time range (time chopping, time cutting). This approach was evaluated and found effective for comprehension of 2D movement visualizations by Amini et al. [ARH*15]. Trajectory semantic enrichment [PSR*13] is also commonly seen via annotating segments, coloring, or incorporating glyphs to encode the primitives or derivatives of the trajectory (e.g., time, space, speed). For encoding only speed and time on 2D paths, Perin et al. [PWPC18] evaluate approaches and make design recommendations.

While these works thoroughly cover the visualization of trajectories and their derivatives, it lacks coverage for internal or external factors affecting objects. Additionally, they do not present a close enough analogue to our target domain. This makes it difficult to determine which of these concepts are relevant or which concepts are missing. Our users need to perform analysis on a multitude of attributes recorded during flight, some of which are abstract. It is not clear in the existing literature how well abstract data can be visualized on trajectories using color.

**Research Question:** Are 2D path color encodings useful for abstract data, moving beyond trajectory primitives or derivatives?

### 2.3. Autonomous Vehicle System Visualization

Despite little academic attention, visualization techniques have been widely adopted for developing and operating autonomous vehicles. For space reasons, here we discuss only extant tools for post-flight analysis of autonomous UAV logs. Among many popular systems for autonomous UAV development, the open-source PX4 [MHP15, PX419d] stands out. PX4 Flight Review [PX419b] is designed to provide an overview of the most critical flight data. It shows a predetermined set of juxtaposed and superimposed [JE12] line charts as well as the flight path in a static 2D or animated 3D view. Relatedly, FlightPilot [PX419a] enables custom selection of attributes to superimpose on a single line chart and PlotJuggler [Fac19] supports multiple line charts through custom juxtaposition and superimposition of attributes. pyFlightAnalysis [Liu19] includes, alongside the line charts, a 3D view of the UAV orientation. However, these alternatives do not show geospatial data. Python libraries [PX419e, Dro19] are available for parsing Ulog files to perform custom analyses. Notable commercial tools include AirData [UAV19] and DJI [DJI19], which have comparable log analysis capabilities to PX4 Flight Review.

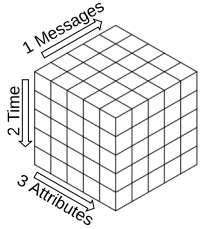

**Figure 3:** *The PX4 ULog data model is a 3D table. 1D: Messages from specific systems, e.g., GPS. 2D: Timestamps for each message. 3D: Attributes for each message.*

We found all of these tools to be serviceable for completing select visualization tasks. However, extant autonomous UAV log visualizations are incomprehensive and juxtaposed charts have little coordination. In addition to this, very little of the state of the art described by geospatial movement data literature is leveraged by in these tools or familiar to their users. Despite the shortcomings of existing tools, they are well integrated into the workflow of our target users. In order for improved designs to be adopted in this domain, they must be easily integratable into these existing workflows.

---

**Research Question:** Can we meet users where they are and integrate with existing workflows but still leverage current research?

---

## 3. Data

In developing DATA COMETS, we used flight logs generated from UAVs using the PX4 autopilot [MHP15, PX419d] flight control software. There are over 3000 different attributes which may be contained in a given log. Some examples include: latitude/longitude coordinates, pitch, roll, yaw, thrust, altitude, velocity, battery remaining, CPU usage, and GPS jamming — to name a few. We collected this data from logs shared publicly by PX4 users [PX419c].

### 3.1. Data Abstraction

Using the data taxonomy presented by Munzner [Mun14], we abstract the flight log data. Each flight log is a spatial temporal multidimensional table. The data is recorded in time and space, giving each datum geo-spatial temporal attributes. These attributes are organized within smaller tables that make up the larger three-dimensional table (fig. 3).

**First Dimension:** The first dimension of the dataset is referred to in the PX4 documentation as system "messages". Each system message contains data relevant for a particular system within the UAV. For example, the system message `vehicle_gps_position` will contain data recorded or used by the UAVs GPS system. The variety and quantity of messages within a flight log will vary depending on the particular UAV it was recorded from. For example, a UAV equipped with a camera gimbal will have messages pertaining to the camera systems but this message would not be present on a UAV without such a system. PX4 handles autonomous navigation and operation in mainly a three-level hierarchy which in turn creates a hierarchy of messages. At the highest level, there is the command that was planned for, or given to, the UAV to execute. This could be an in-flight command to fly to a specific coordinate. The next level would be how the UAV estimates it will execute the command. In other words, the UAV will plan its path to the given coordinate. The lowest level is how the UAV actually executed the given command — the recorded flight path to the coordinate.

**Second Dimension:** All data in the flight log is temporal, making our second dimension time. Each message contains a timestamp. The frequency with which the data was recorded varies depending on the system the message is from. Within the same UAV, gyroscope sensors will record data at a higher frequency than, say, sensors gathering temperature information. This could range from a few times per flight to several hundred times per second.

**Third Dimension:** The third dimension is made up of the attributes each message contains. Messages with spatial information will contain data for spatial position (lat, lon) as well as relative position (x, y) and orientation (pitch, roll, yaw). Attribute types are a mix of categorical and quantitative with mainly a sequential ordering. However, there are some attributes with diverging and cyclical ordering. Cyclical data includes spatial orientation such as roll, pitch, and yaw. Diverging data includes mostly hardware systems such as active control surfaces (i.e. rudders and flaps) or propulsion.

## 4. Grounded Evaluation Design Process

UAV and autonomous vehicle visualization is not well covered by visualization literature. However, relevant topics such as geospatial visualization of movement data are extensively discussed. Under many existing development models, literature on these relevant topics is adequate for attempting task analysis and abstraction with domain experts. Even with our knowledge of related topic areas, we believed there was much to be learned in our target domain before continuing to later development steps. To fill this gap, we decided to take a grounded evaluation route and begin our development cycle with evaluation. However, we wanted to refine what was presented by Isenberg et al. [IZCC08] with a development model that better reflects the nuanced changes that occur with each successive iteration of the cycle. We also believe the model would benefit from additional clarity of the processes that should be done at each step. By understanding the connections between many development cycle models (fig. 2), we determined that the stages in each iteration of our cycle should consist of *Evaluate*, *Task Analysis*, *Design*, and *Implement*. We can also follow these connections to allow us to draw from the guidance detailed in other models for related steps.

We present a model for grounded evaluation that starts the cycle with no direct user involvement and progressively incorporates it over every iteration. This development cycle can be best described as a spiral, where achieving user goals is at the absolute center (fig. 4). We describe user goals as the functions, tasks, and needs a user could have for a given visualization tool or system. Each iteration of the cycle (or revolution of the spiral) represents a different phase in development. As development progresses, the breadth of user involvement increases in each subsequent iteration as a result of conducting more interviews, further evaluations, and eventually deploying the tool to a larger user base. In addition, each iteration of the cycle elapses quicker as you progressively get closer to fulfilling user goals. Iterations where you are implementing the majority of a visualization's features will take longer than post-deployment iterations when only small tweaks are needed. If this is not true, it may indicate that the current project has plateaued. To avoid feature creep and other development pitfalls it may be best to begin a new project and a new cycle. We formulate that this development cycle consists of three distinct phases: early, middle, and late.

**Early Phase:** These iterations will include literature review as well as the initial evaluation of domain problems, users, data, and existing visualizations or solutions. The goal of this phase is to achieve a well-informed understanding of the domain as it exists. This will help researchers begin to establish the context of their work's intended use. An initial task analysis can also be completed by extracting abstractions from previous literature and inferences of user goals from the evaluation material. During this phase, researchers should also be able to create prototypes and sketches with information learned during the evaluation and task analysis.

**Middle Phase:** In this phase, researchers recruit and interview domain experts in order to validate their early phase task analysis, designs, and implementations. In doing so, they will also generate and refine task abstractions, design requirements, and more complete implementations of the visualization — incorporating the feedback received from experts. With the help of the domain experts, the researchers should also identify and plan what is required for a minimum viable product. This includes planning how the visualization will be accessed, how users will input data, and other features not directly related to visualization. Planning these aspects early will help ensure a visualization can be easily deployed and integrated into the workflow of the target users.

**Late Phase:** At this phase, a ready-for-use visualization tool will have been created and deployed to the target users. The broader involvement of active users will continue to provide feedback and suggestions which can be implemented to further refine the design. Researchers should engage with users to fix bugs, implement quality of life features, plan for future functionality, and ensure the visualization generalizes to the larger user base. Revolutions during this phase can also inspire redesigns or new projects.

### 4.1. Initial Evaluation

Following our development cycle, we produced our initial task analysis, design, and prototype. We started with an evaluation of existing domain tools, tasks, and problems.

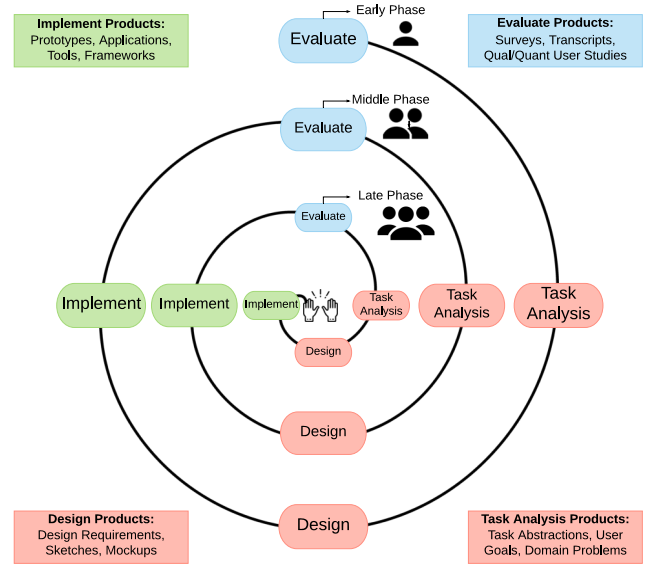

**Figure 4:** *Spiral representation of our semi-cyclical development process. The process starts with Evaluation in the Early Phase and iterates infinitely towards user goals in the center. Total user involvement (represented by the user icons) should grow gradually with each iteration, eventually reaching the Middle phase (domain expert interviews) and Late phase (deployment and end user feedback). Expected products of each development stage are shown.*

#### 4.1.1. Existing Domain Tools

We reviewed and critiqued many of the visualization tools currently used for flight log analysis. During our initial evaluation of PX4 documentation and forum communication, we searched for tools using combinations of the terms "UAV Flight Log Visualization" and followed referrals from domain expert interviews. We selected tools that focused on visualizing post-flight UAV log data and were open-source or well documented. Commercial UAV products were excluded if they did not provide demos, documentation, or screenshots to assess. In total, we reviewed ten tools.

We then organized tools by the visualizations and basic interactions used and according to their general use case. They fell into three categories: flight maps, overviews, and plotting tools. Flight map tools focus mainly on visualizing the flight path the UAV took. This is done by overlaying the flight path over a map, in 2D or 3D, to show the geographic location of the flight or an arbitrary coordinate space. Overview tools present users with curated plots meant to give a high-level overview of the flight data. This includes smaller flight path encodings and time series line charts plotting a select number of hierarchical or related attributes together. There are also a number of plotting tools specifically made to read PX4 flight logs that allow users to select data to visualize with line charts. Several tools we reviewed are detailed in section 2.3 and the review is included as supplemental material at osf.io

#### 4.1.2. Domain Tasks and Problems

Next, we passively observed our target users by reviewing PX4 documentation, Slack communication, GitHub issues, and forum posts. The motivation behind this was to better understand what issues our intended users were experiencing and how they could be resolved. We observed a number of issues related to: testing new development builds, usage and configuration, and debugging errors or malfunctions that happened during a flight. We found that it is common for flight logs to be shared along with screenshots and observations made from analysis tools to assist in troubleshooting.

### 4.1.3. Takeaways

With the results from this evaluation, we were able to formulate a rough initial task abstraction and a list of potential design requirements to build a prototype. We saw the benefit of all three categories of flight log analysis tools and wanted to incorporate elements of each to minimize the need to jump from one tool to another. Users often used flight maps to pinpoint where and when issues happened during the flight, such as a crash or a failed command. They would then refer to this timestamp when using subsequent tools. To aid this, we wanted to enable users to filter and interact with a flight map and highlight selected areas on related attribute line charts. We also noticed that existing tools mainly treated data either as geospatial or temporal. This neglects the highly interrelated nature of this geospatial, temporal, and sometimes abstract data. Finally, many flight map tools play animations in order to visualize the movement of the UAV over time. Waiting for an animation to progress disconnects users from their data. To address this we wanted to use interaction instead of animation to visualize the UAV movement. We implemented these design ideas in a basic visualization prototype using a small subset of data available in a given flight log. This concluded our early phase development and allowed us to move to the next phase of the development cycle.

### 4.2. Task Analysis

We began our middle phase by reaching out to domain experts to participate in semi-structured interviews. We focused on recruiting active and contributing members of the PX4 community. We identified these individuals by reviewing GitHub commits, observing public forum communication, and attending local related meetups. Experts that were recruited consisted of three developers and three pilots from the PX4 community and a UAV startup. All of the participants were representative of our target users: having substantial experience with UAV development, operation, and flight log analysis. During the first half of the interview we asked participants questions about their UAV usage, log analysis frequency, general workflow, and challenges or problems they experienced with existing tools. For the second half, participants interacted with our prototype while providing feedback and answering questions similar to those asked before. We noticed that participants would often refer back to the first half of the interview with new insights and comments after having used our prototype. For example, we asked subjects where they thought existing tools were inadequate. Many of the thoughtful responses to this question only came after subjects had used our prototype. We found that with an interactive visualization in hand users were able to give more specific answers to our questions while engaging in an organic discussion around flight log analysis. We watched our target users utilize our initial design choices in a way that demonstrated their potential, or in other cases identified where they need to be improved or abandoned.

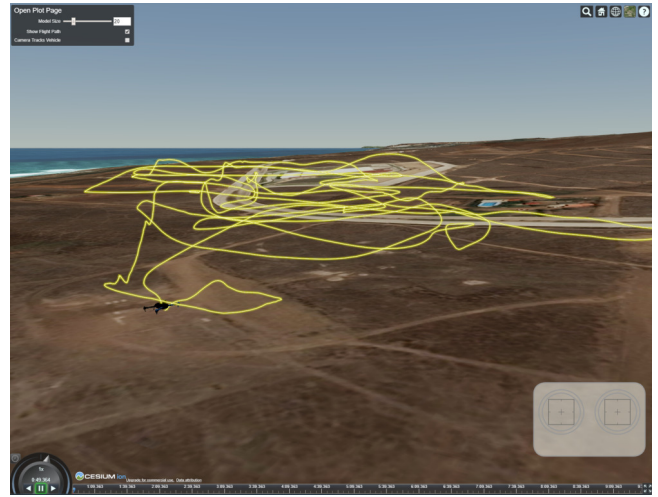

**Figure 5:** *3D flight path visualization from Flight Review [PX419b]. Users can select play to watch 3D model of a drone animate along the flight path. The same flight log can be seen with* DATA COMETS *in fig. 6.*

We recorded these interviews, transcribed them, and performed open coding as described by Vollstedt [VR19]. The coding, performed by two researchers, began by first organizing the semi-structured responses back to the respective broader questions we had prepared. We then summarized sections of responses with similar sentiments, unique insights, or unexpected answers. Finally, we created several codes that encapsulated common themes that appeared in these summaries. Codes included labels such as collecting observations, anomaly or error detection, and data comparison.

### 4.3. Task Abstraction

After performing open coding on the semi-structured interview transcripts, we use the Brehmer et al. [BM13] task typology to create the following abstract visualization tasks [LTM18].

1. **Overview of flight data** (Explore→Summarize): When opening a flight log users will want to first establish context for the flight. Users need to see where the flight happened, what the path and trajectory was, and at a high level how does the data look. This includes quickly checking to make sure the UAV executed most or all of its expected functions correctly.
2. **Discover events or anomalies** (Explore→Summarize): After an overview the user will want to look closer to find any parts of the flight that may have gone wrong. This includes answering questions such a: was there a crash, did the UAV fail any commands, or did the UAV act unexpectedly.
3. **Locate when and where** (Locate→Identify): If anomalies or errors are detected, the user will want to focus their analysis on those problem areas. This includes where the UAV was in its flight and at what time did events occur.
4. **Identify source(s) or cause of error** (Locate→Identify): After the user has found the times and locations of events, they will begin to start looking through the data to find a cause. This includes checking the various UAV systems to find problems in the recorded data. Additionally, it is important to see the data recorded directly prior to and after an identified issue.

5. **Compare data from many sources** (Lookup→Compare): An error can cause issues in several parts of the UAV. Due to message hierarchies and inter-system dependencies, an error in one system is likely to cause an error in several. Users need to check and compare several sources simultaneously to view hierarchies and dependencies in order to successfully diagnose a problem.

## 5. Design Requirements

Using these task abstractions we formulated a series of design requirements DATA COMETS will need to meet user goals. These requirements can be summarized into broader categories as follows.

**DR 1: Visualize Geospatial Data** — The geospatial attributes of the data should be encoded to show users where this data occurred in space. This is not only true for explicitly geospatial data types, such as latitude and longitude coordinates, but potentially all of the recorded data. The data is recorded in space and time, where it happened could be just as important as when it happened. Users should also be able to contextualize the environment where the flight occurred, e.g. in an open field or a city.

**DR 2: Visualize Temporal Data** — All of the flight data is organized by when in the flight it occurred or when it was supposed to occur. This data should be encoded in such a way that lets users see how it changes over time.

**DR 3: Provide Overview** — Users need contextualized views of the data that allow them to begin deeper analysis. High-level or important attributes should be encoded and displayed to provide this context. Users should also be able to visualize UAV movement.

**DR 4: Provide Detail** — Deep analysis of the data should be enabled by providing additional details or focused views on demand. This includes being able to filter the data shown both spatially and temporally. Users should be able to compare specific values and changes across all attribute data. All attributes present in the flight data should be available for analysis and all of the attributes dimensions, e.g., value, space, and time, should be visualized.

**DR 5: Enable Comparison** — Data from many sources and across system messages will need to be analyzed for successful analysis. Attributes should be displayed in a manner that enables them to be easily compared across all of their dimensions: value, space, and time. Users should be able to select which attributes they want to compare using any of the visualizations provided. These views of the data should allow users to see how changes in one or more attribute lead to changes in other attributes.

## 6. Visualization Design

DATA COMETS is an open-source UAV flight log analysis tool built to support PX4 system data. The tool was built with a web stack utilising a Flask-based [Pro19] Python back end to parse the data and a JavaScript/HTML/CSS front end. The front end was developed using D3.js [BOH11], Leaflet.js [Lea19], Simplify.js [Aga19], and Materialize CSS [AW19]. DATA COMETS aims to enable users to effectively analyze flight logs to verify system correctness, identify system errors, and debug system failures. To accomplish this DATA COMETS is comprised of four components: flight map, attributes tree, timeline, and interaction, that each play a vital role in addressing our design requirements (fig. 6).

### 6.1. Flight Map

The flight map is the most prominent view in the tool (fig. 6: A). This gives ample screen real estate to allow for overlays, encodings, and easier user interactions. It uses an ArcGIS high resolution satellite image based projection [Arc19] that is centered to a reference latitude and longitude (lat/lon) coordinate present in the flight data. We chose to use a satellite image based map, over a more abstract street or topology based one, to give more context to the user. Being able to see certain landmarks, terrain features, and buildings can help users establish where the flight happened and under what conditions. In some cases this could allow users to identify which flight they are looking at if they have several flights over many locations to analyze. This can also help users determine the flight condition they should be expecting to see. A flight in a city — in close proximity to buildings — is more likely to show poorer signal quality versus a flight in a rural area. Users can also pan and zoom on the map to view different sections of the flight path that might not be visible in the initial view. Zooming in allows users to see smaller details of the flight path that could have been obfuscated when zoomed out. This is mostly noticed with smaller horizontal movements such as when the drone is hovering, performing subtle maneuvers, or taking off and landing. At high zoom levels, the background changes to a solid gray grid instead of displaying a low-resolution pixelated map. The gray background is also displayed when no satellite image is available, including when there is no internet connection to query map tiles. This first impression of the flight location supports *DR 1: Visualize Geospatial Data* and *DR 3: Provide Overview*.

The recorded flight path is overlaid onto the map to show the route the UAV took (fig. 6: E). This is encoded using a series of paths starting at the current lat lon coordinate and ending at the next. Encoding in this manner results in a visualization of the full flight path with individual paths representing the time and distance between records. We leverage this effect to encode other attributes onto the flight path using a colorblind-safe palette with equal lightness steps ▬▬▬▬▬▬ [Ais19]. For example, we can encode the UAV's velocity onto the flight path as a color map by aligning the timestamps of the data and encoding the corresponding paths with a respective color. This allows users to view where geospatially attributes occurred. In the example where we encode velocity (fig. 1), this will show how the UAV slows down to change directions, and potentially could show how it failed to do so causing it miss its target trajectory. If velocity was encoded only as a line chart, this geospatial aspect would not be immediately understandable. A line chart does not show events such as a change of direction, it only would show the aforementioned reduction in velocity or lack thereof. Encoding attributes on the flight path is also of particular importance for attributes that are influenced by location such as GPS jamming. In the event that a particular location exhibits high levels of EM/RF interference, it would be visualized by the color of the paths in that location. This allows users to quickly see areas of the flight that had trouble with similar attributes.

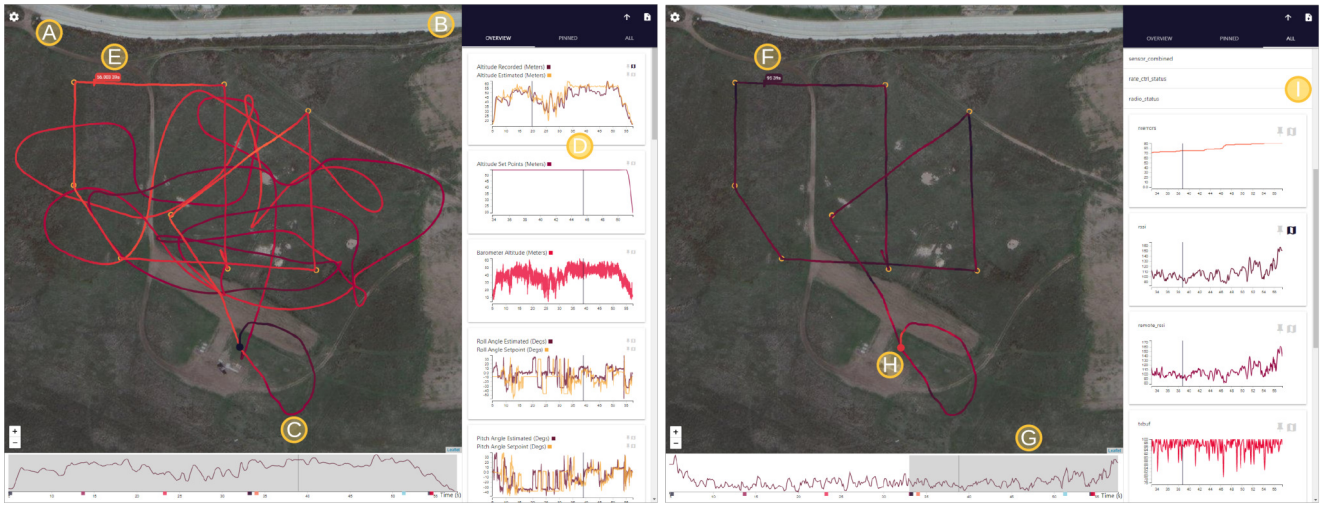

**Figure 6:** DATA COMETS *visualizing the same flight log scene in fig. 5. All data is selected with the attribute tree displaying Overview tab (left). Data is filtered to only show data from the beginning of autonomous flight with the attribute tree displaying All tab (right). Demonstrated: (A): satellite image based map, (B): attributes tree panel displaying Overview tab, (C): timeline with all data selected, (D): line chart hover interaction, (E): flight path encoding selected attribute (altitude) with color, (F): filtered flight path with hover interaction, (G): filtered timeline, (H): glyph representing the current drone position from data selected encoding selected attribute (RSSI) with color, (I): flight message drop down elements in All tab.*

We chose to use color for the additional data dimension because of its capability to encode abstract attributes [Mun14] while not introducing excess visual clutter. Color has been used effectively in a wide range of applications linking spatial and non-spatial data [CWK*07, GRW*00, OSS*17]. Moreover, it is straightforward for our users to comprehend as it is used in their existing tools.

Alternative encodings using tick-marks and segment width [PWPC18] are not robust enough to encode many of the attributes found in a typical flight log. Tick marks can indicate the distance between points of a trajectory (primitives) which can be used to infer the time and speed between segments (derivatives), but ticks are challenging to use for abstract data. We explored segment width encoding in an early prototype as an optional addition to color, however this feature did not evoke a strong response from users. Moreover, width encoding is not suitable for categorical attributes in flight logs, e.g., operating modes and whether sensors are on/off. Width encoding also adds occlusion when trajectory segments overlap or are tightly spaced — commonly the case for UAV trajectories. Reducing the path opacity is not an ideal solution as the satellite image background will influence path color perception in hard to control ways. Overlapping paths are discussed as a potential limitation of these encoding methods in Perin et al. [PWPC18], but this has not yet been evaluated.

These encodings contribute to supporting *DR 1: Visualize Geospatial Data*, *DR 2: Visualize Temporal Data*, *DR 3: Provide Overview*, *DR 4: Provide Detail*, and *DR 5: Enable Comparison*.

By using a settings window, users can also choose to encode other map related info, such as filtered sections, estimated flight path, and set points. The filtered portion of the flight path on the map is optionally shown as a dashed line ▪ ▪ ▪ for focus plus context. The flight set points are the targets the UAV was instructed to fly to in a given order. The estimated flight path was how the UAV system anticipated it would get to each set point. These along with the recorded flight path account for a message hierarchy. The set points are encoded onto the map as donuts ⭕ (fig. 6: E) placed at the appropriate lat lon coordinates. Estimated flight path is encoded as a semi-translucent single colored path ▬ (fig. 1) rendered underneath the recorded flight path. A user will be checking these attributes to see if they line up with each other as expected or if they deviate. These optional encodings give users additional options to visually analyze their data and support *DR 1: Visualize Geospatial Data*, *DR 3: Provide Overview*, and *DR 5: Enable Comparison*.

### 6.2. Attributes Tree

DATA COMETS second most prominent view is a panel that contains line charts and details about all flight data attributes (fig. 6: B). This panel uses a tree like structure that can be navigated with a series of UI tabs and drop down components. The tabs — overview, all, and pinned — organize the data with different views that each offer the users a different way to analyze their data. Each of these tabs will display attributes as line charts and or their value if it is constant. On each line chart, there are UI controls users can click to encode that attribute on the flight path 🗺 or to pin the chart and add it to the *pinned* tab 📌. Attributes from messages with high record frequencies (60+ records/s) ran up against the performance limits of using svg rendering. To increase the interaction and rendering performance for these and all attributes, line charts used simplify.js [Aga19] to reduce the number of points in the svg path while maintaining the original path shape.

*Overview:* The *Overview* tab is the default view of the tree panel and provides users with pre-configured line charts of important attributes (fig. 6: D). These are the attributes that users often look to check first when viewing flight data and can give context to the flight performance or a starting point for what data to look at next. The visualizations are configured and ordered to allow users to easily view and compare hierarchical and related attributes. When appropriate and where hierarchical attributes share a common scale, attributes are plot together on the same visualization. For example, data such as altitude, roll, pitch, and yaw are visualized on a chart with both the recorded and estimated attributes. Furthermore, relevant attributes are configured to be visualized inline with each other and organized so that associated attributes are close to each other. Data related to the UAV's battery levels are placed near each other and so on. Plotting related attributes with or near each other lets user check to make sure the data lines up or changes as it should. Line charts encoding altitude set points, recorded, and estimated should roughly all match if the flight had no errors. This curated view is something users are familiar with from existing tools used to analyze flight logs. We aimed to improve this view users were already familiar with by avoiding some of the bad practices found in other tools. This includes: reducing visual clutter, only plotting attributes together when they share a common scale, and enabling interaction and details on demand. With all these elements the overview tab supports *DR 2: Visualize Temporal Data*, *DR 3: Provide Overview*, and *DR 5: Enable Comparison*.

*All:* From our interviews with domain experts we learned that, while tools that offer curated charts are great for providing overview they often do not show all the data they need to see. Given this users will use other tools that allow them to plot arbitrary attributes, or download the data and create the visualizations themselves. To help easier and deeper analysis, the *All* tab contains all the data present in the flight log (fig. 6: I). The data is organized by a series of drop down elements with one drop down for each message in the log. When the drop down is clicked it expands to show all of the attributes within that message. Creating pre-configured visualizations for all the possible attributes and messages that may appear in the flight log is not feasible. There are 111 different messages and 1,300+ attributes and this number is expanding with every new version of PX4. Aside from the sheer numbers, metadata for units, descriptions, and other details do not exist within the flight logs at this point. Some of this information can be learned through the documentation and source code, but only for a small subset of attributes. To solve this, plots are generated automatically when the flight data is loaded in and use the data available to create the axes and titles. If an attribute has a constant value throughout the flight, a line chart is not used and instead the value is displayed at the bottom of the drop down. This gives users who want to dig deeper into the flight data the ability to easily see attributes that are less commonly checked or are more esoteric. This also allows DATA COMETS to be extend to include new messages and attributes as they are added. Giving the users an organized view of all the flight data available supports *DR 2: Visualize Temporal Data*, *DR 4: Provide Detail*, and *DR 5: Enable Comparison*.

*Pinned:* The *Pinned* tab gives users the ability to select which attributes are shown in the tab (fig. 7). This supports users who will want to dig deeper into the flight logs to analysis attributes less commonly looked at, or that are newly implemented so they would not be included in an overview. These users often will download the data and create these line chart visualizations themselves in order to look at these specific attributes. With the *Pinned* tab they can essentially create these line charts within the tool and display them in the order they desire to enable comparisons amongst several attributes. The view is initially blank and will display the attributes that users have selected with the pin icon found on each chart. The chart will then be shown in the view with the title changed to include the message the attribute belongs to as well as the attribute name. We decided to add the functionality after seeing just how many different messages and attributes there could be in the *All* tab. If a user wanted to compare attributes from different messages or even within the same message, it was often the case that the charts were simply too far apart to appear in the view simultaneously. With the pinning functionality users can add the attributes they want to a dedicated view where they can be seen more comfortably. This allows users to compare attributes across any message that may be related, and also lets them review attributes that are a part of a hierarchy but not shown in the *Overview* tab. The utility the *Pinned* tab provides users to control which attributes to display supports *DR 2: Visualize Temporal Data*, *DR 4: Provide Detail*, and *DR 5: Enable Comparison*.

### 6.3. Timeline

The third and final view of DATA COMETS is a timeline that spans the width of the map at the bottom of the page (fig. 6: C). The timeline serves mainly as a UI component for interaction, but it encodes some flight data as well. Aside from displaying the duration of the flight in seconds or minutes, the timeline also displays a static line chart of the currently selected attribute (the attribute that is encoded on the map). As longer flights (greater than one hour) become more commonplace, this view should be updated to allow users to zoom in for finer control. The timeline also serves as a place to encode certain categorical or miscellaneous attributes such as flight modes and in flight messages. PX4 supports 22 different flight modes that detail the operational state of the UAV. This includes modes such as manual flight, begin auto mission, return to land, and so on. The flight modes are categorically colored and encoded on the timeline as squares just bellow the y axis. The timeline plays a central role for user interactions as well as severing as a place to display a static view of some attributes. In doing this it supports *DR 1: Visualize Geospatial Data*, *DR 2: Visualize Temporal Data*, *DR 3: Provide Overview*, *DR 4: Provide Detail*, and *DR 5: Enable Comparison*.

### 6.4. Interaction

DATA COMETS enables interactions that enable users to more effectively analyze their data. As previously discussed, the timeline serves as an important area for user interaction. Users are able to use brushing to filter the window of time shown on other parts of the visualization. This is represented by a semi-translucent light grey rectangle — or window — overlaid on top of the timeline (fig. 6: G). The user can change the size of this window by dragging either side left or right, as well as move the position of the window by clicking and dragging it along the timeline. Finally the user can draw a new window by clicking on the timeline outside of the current window and dragging. When the time line is brushed the flight path and line chart visualizations both change to only show the range of data within the timestamps of the window. This allows the user to narrow down their focus to certain parts of the data. The line charts zoom in to let users more clearly view the features of the data within the desired window. For the flight path this can also help alleviate visual clutter that can occur with longer flights with many overlapping paths.

Additionally, users can manipulate the window to animate the flight path, either from start to finish or within a certain time window. Being able to interact and re-create the flight path allows users to see what the UAV was doing during the flight in a way that could be lost in a static view [ARH*15]. E.g., the flight path could appear to hit all of the set points, but it would not be clear in what order or in which direction this occurred without playing back the flight. DATA COMETS also supports linking between all of the views. Hovering over the line charts will display a long thin rectangle at the time stamp being hovered as well as the value (fig. 6: D). A similar mark will also appear on the timeline, and the corresponding flight path segment will enlarge to indicate its position on the map (fig. 6: E,C). The same effect will occur when users hover over the flight path segments (fig. 6: F) and the timeline flight mode squares. These interactions are a key way users can analyze and consume their data and improves the usability of all of the views. As a result they support all of our design requirements: *DR 1: Visualize Geospatial Data*, *DR 2: Visualize Temporal Data*, *DR 3: Provide Overview*, *DR 4: Provide Detail*, and *DR 5: Enable Comparison*.

## 7. Evaluation

After completing our middle phase design cycle iterations, DATA COMETS was ready for public use. From speaking with domain experts during the middle phase, we learned that the PX4 community would favor having easy access to the tool online while also respecting privacy. Having access to a version that could be run locally was also desirable as it would offer better performance and increased privacy. To address these needs, DATA COMETS was prepared with a hosted version that never saved uploaded data, and a public repository with instructions on how to run DATA COMETS locally. We announced and launched DATA COMETS by creating a post on the PX4 discussion forum, as well as making an announcement on the PX4 Slack channel. Using these channels we could interact with, observe, interview, and survey our target users in order to conduct an evaluation of user experience as described by Lam et al. [LBI*12]. For more direct feedback we interviewed nine users during the middle and late phases of development. These users all came from the PX4 community in some capacity, either as members of the PX4 team or users of the PX4 software in industry. For the late phase evaluation we recruited the domain experts who we initially interviewed for the task analysis, as well as one new active user who had been engaging with DATA COMETS. Users had been using the visualization tools available or that they had created themselves perform visual analysis of their UAV data. Before the interview, users had either been using DATA COMETS after reading the documentation or were given a quick demo of the tool and its features. We asked participants about their experience using the tool, its fit in their workflow, how it compared to other existing tools, and what features they would like to see improved or added. The transcripts from the interviews were reviewed by two researchers to identify positive and negative sentiments that either support our design or indicate required changes.

### 7.1. Results

#### 7.1.1. User Reception

The DATA COMETS launch announcement was met with a warm response and excitement from many in the PX4 community. This include some of the largest PX4 contributors as well as PX4's founders. Since our official launch August 12th, 2019 to March 4th, 2020 DATA COMETS saw 266 users launching 452 sessions. This included users from a UAV based company who are looking to integrate DATA COMETS into their analysis workflow. In addition to this, shortly after the announcement DATA COMETS was added to the flight log analysis page of official PX4 documentation.

#### 7.1.2. User Feedback

Users commented on DATA COMETS ability to provide a complete overview of flight data. A test pilot commented how the map "jumped out" at them. They appreciated that the satellite imagery was up to date and that it looked like where they were flying so they had immediate context on where they were looking. Users also liked having the map and line chart views connected and displayed in tandem. A user noted finding when and where an issue happened in the flight was "comfy" as they did not have to switch between multiple tools to troubleshoot the problem anymore. Encoding attribute data on the flight path was also an appreciated novel addition. One user commented how it was useful to "at a glance visually see where you could zoom in." This user also detailed how this feature "gives you an awareness that is hard to get from graphs side by side" and gave the example of checking for network interference.

Users were also able to conduct deeper analysis of the data with DATA COMETS after completing an initial overview. This was primarily supported by the timeline filtering function. Users often mentioned liking being able to "zoom in" both on areas of the map and time intervals in the line charts. This interaction can also be used to visualize the movement of a flight. Users appreciated this vs. the more common animation approach, which was described by one user as "cumbersome" — supporting Amini et al.'s observations [ARH*15]. In addition, users appreciated all the charts being linked through hover interactions as it allowed them to compare many attributes at the relevant time point and pinpoint problems.

Lastly, many users made it a point to mention that DATA COMETS was easy to use and user-friendly. These included comments such as, "I was just able to use it right off the bat. And I think it's rather intuitive."; " it's fairly easy to interpret...it's user friendly...everything is just so easy...the information is so simple to read."; and "with this tool, I'm able to get grasp quite easily on what's happening in a flight, and I have a better idea of what's going on and I can ask better questions to the team with this tool."

In addition to validating feedback, users also provided constructive feedback we could utilize for late phase development. This included bug fixes to support logs from different versions of PX4, missing data, and performance improvements. Users also expressed their interest for new features, such as an expandable chart view, 3D view, and visualizations for UAV orientation.

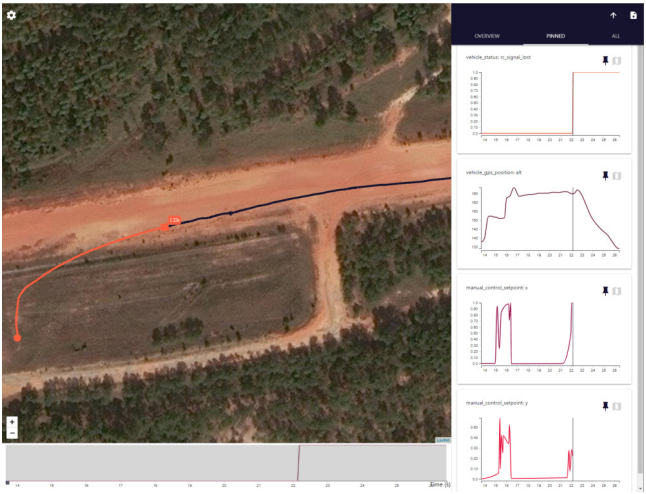

**Figure 7:** *Flight log from a UAV that experienced a malfunction visualized with* DATA COMETS. *UAV lost radio control (RC) connection and entered a fail safe landing mode. Using the pin tab to display the relevant data we can see the exact moment rc was lost and how other attributes were effected by this. Encoding the RC connection attribute onto the flight path shows exactly where the malfunction happened geospatially.*

### 7.2. Usage Scenario

To see if DATA COMETS could be used to analyze and answer issues with flights, we looked back to posts from the PX4 forum and GitHub. In one post [Hur19] a user described that their UAV was entering a fail-safe mode and immediately landing after radio control (RC) is lost instead of finishing its programmed mission. The community identified that the UAV did not have any set points or commands from a mission to execute, and after RC was lost, the UAV resorted to landing and this is the expected behavior. We can see this happening clearly when analyzing the log with DATA COMETS. Figure 7 shows this flight log visualized with DATA COMETS using the pinning functionality to isolate the attributes relevant to the issue. RC signal lost, a binary true or false attribute, is encoded on the flight path and shows when and where the UAV was when the signal was lost. Looking through the *All* tab, we can see the log message for position set points used for auto pilot was empty. The only set points recorded were the manual control set points visualized in the attributes tree in fig. 7. We can see that immediately after RC was lost, there were no more manual control set points, as there was no connection back to the controller. With no mission set points to follow, the UAV entered a fail-safe mode, and began its descent for landing. In addition to visualizing this flight statically, we can use DATA COMETS timeline brushing interaction feature to visualize the UAV movement. By brushing the time window from start to end, we can see how the UAV flew from the left of the map to the right, and for a few seconds before losing RC it was hovering in place. In existing tools, this small nuance would have only been noticed had a user been present during the flight or used a separate flight path visualization tool to watch an animation of the flight.

## 8. Research Questions Revisited

### 8.1. A Grounded Evaluation Model for Visualization Development — Theory and Practice

Grounded evaluation proved to be especially effective for this study and the development of DATA COMETS. The knowledge gained from our initial evaluation of the domain was critical to the later phases of development. By the time we recruited and spoke with domain experts, we had already implemented a working prototype and had become familiar with their common problems and practices. We noticed that having our early prototype ready helped raise interest in what we were doing and made the recruitment process easier. Prior to interviews, having an understanding of the data, tools, and general work flow allowed domain experts to easily convey their tasks without needing to provide background information. Without our initial evaluations, we would not have been capable of implementing such a complete prototype so soon into development.

Additionally, the spiral development model not only helped organize the research and development of DATA COMETS, but it allowed us to go from complete outsiders to participating members of the PX4 community. Our initial evaluation of the domain was central to this, as we were able to identify domain experts who could help us during the later phases of development. These experts not only aided during task analysis and evaluation, but also with deepening our connection to the community and allowing us to continue broadening user involvement. This ultimately culminated in a tool that we could test close to the context it was designed for — a central goal of grounded evaluation as described by [IZCC08].

### 8.2. 2D Path Color Encodings for Abstract Data

We knew from the exiting literature that trajectory semantic enrichment, such as time or space coloring, would be a suitable method for visualizing UAV data. However, encoding the path with color had primarily been explored for encoding primitives or derivatives of the trajectory itself (time, space, speed, acceleration). None the less, we found this technique to be robust enough to effectively encode a diverse range of attributes — even those not directly related to the trajectory. This included, binary data such as RC lost fig. 7, categorical data such as flight modes, continuous data such as battery level, and diverging data such as actuator position. We found our users quickly became comfortable interacting with and reading these encodings, and enjoyed being able to map any of the data available in this manner. This technique was further aided through linking between the flight path and the line charts as well as the timeline filtering operation (time chopping). Both of these interactions helped users to view the attributes they wanted to see at the space or time slot they were looking for.

### 8.3. Meeting Users Where They Are with Existing Workflows

Our users all had their own workflows for visual analysis of their data. While they all expressed displeasure with the tools that were currently available, that did not mean they would adopt new designs readily. Many concepts presented in related visualization literature would potentially have been foreign for many users. Thoroughly understanding the tools our users were already familiar with was critical to insuring our design would integrate easily into their work flow and mental models. With this information we identified the concepts from literature that could map easily to how existing tools were being used. As a result, DATA COMETS can be easily adopted by any PX4 user and was found to be user friendly — requiring little instruction to use.

## 9. Limitations and Future Work

As we are still in the early stages of DATA COMETS' public deployment, it is difficult to assess long term implications. The tool has been adopted by the PX4 community as a new method for analyzing flight logs. However it is still too early to tell whether or not DATA COMETS will become part of the daily work flow of our users or if its initial positive reception was just due to novelty. We are eager to observe DATA COMETS usage within the community by measuring long-term feedback, applications, as well as web analytics such as new and returning users, session duration, and downloads. We are also interested in quantitatively evaluating aspects of DATA COMETS design compared to existing tools. This would require controlled studies using purposefully recorded flight logs with labeled errors, malfunctions, and anomalies. In addition to this we are still iterating through our late phase development cycles, gathering user feedback to plan, implement, and tweak features to better meet user goals. Future versions of DATA COMETS will explore ways to better integrate 3D views as well as expanding visualization customizability.

## 10. Conclusion

We presented a design study for autonomous UAV system visual analysis. Using a grounded evaluation development model, we iterated through our phases of development, beginning by evaluating the domain as it existed, identifying areas for improvement, and implementing prototypes. We then recruited and interviewed domain experts to further understand their work and tasks, iteratively improving our design. Finally, we deploy and present the design and implementation of DATA COMETS, a visual analysis tool for UAV developers and operators. DATA COMETS utilizes geospatial path encodings to visualize not only when data happened but where, as well as interaction and filtering techniques to understand the movement of the data. We evaluate our design with domain experts and users directly and confirmed that our design could be used to effectively validate, identify, and diagnose correct or incorrect system behavior. DATA COMETS has been adopted as a tool for flight log analysis by the PX4 community and will continue to improve with each new iteration of our development cycle.

## 11. Acknowledgments

We thank the U.S. ONR for support under OTA N000141890001, members of the Visualization Lab at Northeastern University, and members of the PX4 community.